\documentclass[10pt, conference, letterpaper]{IEEEtran}
\IEEEoverridecommandlockouts

\usepackage{cite}
\usepackage{amsmath,amssymb,amsfonts}
\usepackage[T1]{fontenc}
\usepackage{algorithm}
\usepackage{algpseudocode}
\usepackage{amsmath}
\usepackage{amsfonts}
\usepackage{float}

\usepackage{textcomp}
\usepackage{svg}
\usepackage{amsmath}
\usepackage{xcolor}
\usepackage{graphicx}

\usepackage{graphics}
\usepackage{float}
\usepackage{cclicenses}
\usepackage{xspace}
\usepackage{subfigure}
\usepackage{makecell}
\usepackage{multirow} 
\usepackage{wrapfig}
\usepackage{enumitem}
\usepackage{booktabs}
\usepackage{url}
\usepackage{graphicx}      % For including images
\usepackage{subfigure}    % For subfigures

\newcommand{\name}{\texttt{AutoMDT}}

\begin{document}
\title{Modular Architecture for High-Performance and Low Overhead Data Transfers}
% \iffalse

% \author{
%     \IEEEauthorblockN{Rasman Mubtasim Swargo\IEEEauthorrefmark{1}, Engin Arslan\IEEEauthorrefmark{2} and Md Arifuzzaman\IEEEauthorrefmark{1}}\\
%     \IEEEauthorblockA{\IEEEauthorrefmark{1} Missouri University of Science and Technology, Rolla, USA
%     \\\{rs75c, marifuzzaman\}@mst.edu}\\
%     \IEEEauthorblockA{\IEEEauthorrefmark{2}Meta, Menlo Park, USA
%     \\enginarslan@meta.com}
% }

\author{\IEEEauthorblockN{Rasman Mubtasim Swargo}
\IEEEauthorblockA{Missouri University of Science and Technology\\
rs75c@mst.edu} 
\and \IEEEauthorblockN{Engin Arslan}
\IEEEauthorblockA{Meta\\ enginarslan@meta.com}
\and \IEEEauthorblockN{Md Arifuzzaman}
\IEEEauthorblockA{Missouri University of Science and Technology\\ marifuzzaman@mst.edu}
}
% \fi
% \author{\IEEEauthorblockN{Anonymous Authors}}
\maketitle              % typeset the header of the contribution

\begin{abstract}
High-performance applications necessitate rapid and dependable transfer of massive datasets across geographically dispersed locations. Traditional file transfer tools often suffer from resource underutilization and instability due to fixed configurations or monolithic optimization methods. We propose~\name, a novel Modular Data Transfer Architecture, to address these issues by employing deep reinforcement learning based agent to simultaneously optimize concurrency levels for read, network, and write operations. This solution incorporates a lightweight network–system simulator, enabling offline training of a Proximal Policy Optimization (PPO) agent in approximately 45 minutes on average, thereby overcoming the impracticality of lengthy online training in production networks. \name~’s modular design decouples I/O and network tasks. This allows the agent to capture complex buffer dynamics precisely and to adapt quickly to changing system and network conditions. Evaluations on production-grade testbeds show that~\name~achieves up to $8X$ faster convergence and $68\%$ reduction in transfer completion times compared to state-of-the-art solutions.
\end{abstract}

\begin{IEEEkeywords}
Data Transfer Optimization, High-Performance Networks, Modular Architecture 
\end{IEEEkeywords}

\maketitle

\section{Introduction}\label{intro}
% From financial systems to scientific research, enormous amounts of data are produced every day. Often, these sites are not located in the same geographical region, and the data must be transferred for simulation, analysis, or experimentation over high-speed networks ~\cite{deelman}. For example, data from the Large Hadron Collider \cite{az6} and the Vera Rubin Observatory \cite{az5} are transferred at tens of gigabits per second (Gbps) between geographically distributed sites at regular intervals. High-speed networks like ESnet provide bandwidths of up to 1000 Gbps. In many cases, this bandwidth is underutilized due to suboptimal data transfer configurations. Widely used file transfer applications, such as FTP, do not employ parallel transfers and thus fail to utilize the maximum available network bandwidth.

Scientific applications, ranging from large-scale computational simulations and machine learning modeling to intricate physical experiments, generate vast amounts of data that must be transferred swiftly and reliably between geographically distributed High-Performance Computing (HPC) clusters~\cite{ligo,ares,combustion,astronomy,des,lsst,atlas,belle2,ameriflux,esgf,esxsnmp}. As science projects are increasingly distributed and collaborative, the massively growing data sizes demand high-speed data transfers to move data between geographically dispersed institutions in a timely manner. Internet2 has upgraded its backbone network bandwidth to 400 Gb/s as the amount of data transferred over its network increases exponentially~\cite{internet2400Gbps}. For example, advancements in high-throughput genome sequencing technology increased output size per single run from around $5$ MB in 2006 to more than $700$ GB in 2024, more than a thousand-fold increase in just 17 years. Consequently, to support large-scale data movements, ESNet has also been testing 400 Gb/s network and terabits per second networks are expected to arrive soon~\cite{esnet1000Gbps}. 

Most studies leverage concurrency (transferring more than one file at a time) to increase utilization of the available bandwidth of these high-speed networks. While increasing concurrency by raising the number of TCP streams can boost throughput, exceeding the optimal number of streams may lead to network congestion, packet loss and end-systems overhead. The optimal solution depends on various factors, such as per-connection bandwidth, background network traffic, and hosts I/O and computing capabilities, all of which are dynamic. Consequently, a fixed configuration prior to the transfer is not effective in addressing these changing conditions. Active probing~\cite{probing} is one approach to finding the optimal TCP stream count; however, frequent probing can itself cause network congestion. Similarly, heuristic and supervised models perform well in specific network environments but struggle to adapt to dynamic situations. Thus, more recent studies approach this problem as an online optimization problem and dynamically tune the value of the parameters.

However, existing solutions follow a monolithic architecture that allocates the same concurrency level to the read, write, and network operations, even though in most production systems these operations require different levels of concurrency, leading to over-subscription of limited resources. Marlin~\cite{marlin} addresses this issue by separating the concurrency levels for read, network, and write operations. However, it treats the problem as multiple single-variable optimizations, which overlooks the intricate dynamics among different components and results in unstable and suboptimal solutions.

% Hasibul et al. \cite{drl} also applied deep reinforcement learning methods. However, they optimize only one parameter and still require 28 hours of online training, making the solution impractical. 

In this study, we introduce~\textit{AutoMDT}, a novel policy-driven deep reinforcement learning (DRL) based optimizer, to jointly predict the optimal concurrency values for read, write, and network operations. To address the long convergence time challenges associated with online DRL training, we designed a simulator that emulates the memory-buffer dynamics of production systems for offline training. The training could be done using the simulator in as little as 45 minutes on average compared to days taken by previous studies~\cite{drl}. In summary, the major contributions of this paper are:
\begin{itemize}
    \item We present a novel DRL based approach to optimize concurrency values for read, network, and write operations. Unlike previous approaches, we jointly optimize all three variables using a single optimizer, thereby learning the complex dynamics among them, resulting in significantly more stable throughput. 
    
    \item To accelerate the \name~agent's learning process, we introduce a network system simulator that emulates dynamics among all relevant parameters. This reduces the training time from approximately 7 days (if done online) to approximately 45 minutes on average.
    
    \item We demonstrate that~\name~can autonomously identify the slowest operation at a faster rate than its predecessors. It reaches the highest network bandwidth utilization up to $8X$ faster and finishes transfers up to $68\%$ faster than state-of-the-art solutions.
\end{itemize}
%\vspace{-2mm}

\section{Related Works}\label{sec:related}
To fully exploit modern gigantic HPC networking infrastructures, several transfer parameters must be optimized, including pipelining~\cite{TCP_Pipeline, farkas2002}, parallelism~\cite{R_Lee01, R_Hacker05, R_Karrer06, R_Lu05}, concurrency~\cite{kosar04, Kosar09, R_Liu10}. These parameters can be tuned at the application layer without the need to change the underlying transfer protocols and can significantly improve the end-to-end data transfer performance. As a result, numerous studies over the years have been working on optimizing different application layer parameters.

% Several application-layer solutions have been proposed to optimize file transfer parameters. Notable techniques include concurrent file transfers~\cite{concurrency}, parallel TCP streams~\cite{parallel}, pipelining~\cite{pipelining}, and using multiple Data Transfer Nodes (DTNs)~\cite{dtn}. The main challenge with these methods is the large search space and the inherently slow evaluation process.

Previous studies for application layer data transfer optimization could be categorized as heuristic solutions~\cite{globusonline,arslan2018big,europar13, hodson2011moving} or historical data modeling~\cite{harp,kettimuthu2014modeling,zulkar-ndm14,arslan2018high,nine2020two}, or online optimization~\cite{rao2016experimental,yun2017data,liu2018toward,yildirim2015application,concurrency,arifuzzaman2023falcontpds,marlin,drl}. Both heuristic and historical modeling use fixed parameter values for the entire transfer duration and cannot adapt to the constantly evolving networking dynamics of the production systems. Fixing values conservatively often leads to underutilization; aggressive values create high system overhead during concurrent transfers. Another key drawback of historical modeling is its reliance on large-scale datasets from various transfer settings using active network probing~\cite{10824943}. It is difficult to collect these data in real-world production networks, as active probing risk causing network congestion due to the additional traffic burden. Also, training data often gets outdated, and we have to recollect and retrain them periodically. To address these concerns, most recent studies use online optimization for adaptive solutions, and our proposed optimizer also falls into this category.

However, all of these solutions are based on monolithic architectures where I/O and network tasks are tightly coupled. FDT~\cite{fast_data_transfer}, mdtmFTP~\cite{The_MDTM_project}, and Marlin~\cite{marlin} move away from this design and separate network and I/O tasks. However, FDT and mdtmFTP rely on manual configuration tuning that lacks adaptability, while Marlin suffers from unstable and suboptimal solutions due to relying on an oversimplified online optimizer. To overcome these limitations, our work introduces a reinforcement learning-based fast and stable modular data transfer architecture named~\name. \emph{In this study, we completely rethink the optimization architecture of Marlin. Here, (i) we investigate the root causes of the stability issues of Marlin, (ii) \name~abandons Marlin's multiple independent single‑variable gradient descent optimizer in favor of a joint
three‑variable reinforcement learning optimization agent for stable solutions and high-performance, and (iii) introduces an
I/O–network dynamics simulator that enables very fast
offline training, avoiding multi‑day online exploration of RL agents.}

%\vspace{-2mm}
\section{Motivation}\label{motivation}
\begin{figure*}[ht]
    \centering
    \includegraphics[width=0.72\textwidth]{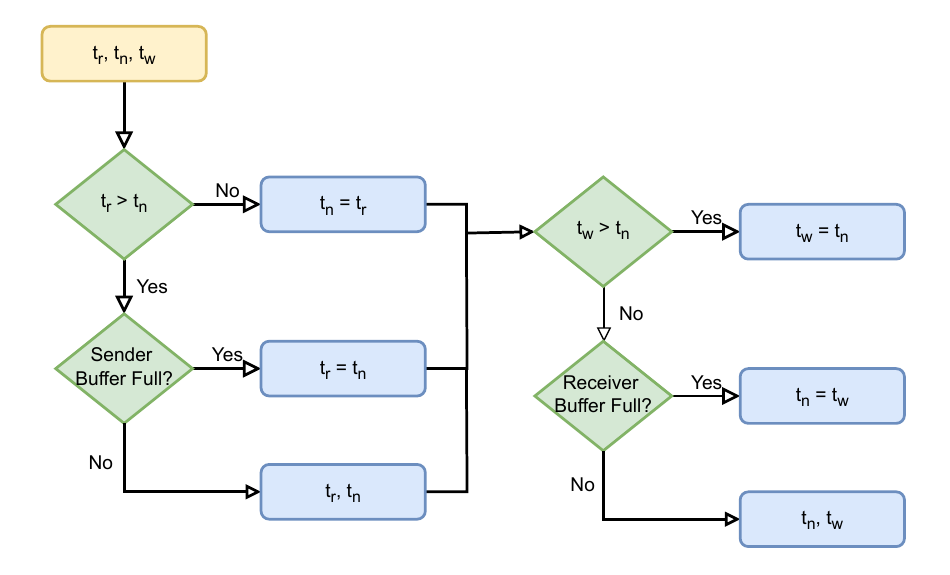}
    \caption{Dynamics of the file transfer process showing the relationship between read, network, and write throughputs.}
    \label{fig:diagram}
\end{figure*}

As modern scientific research networks have links up to $1000$ Gbps, we need many concurrent connections to fully utilize these gigantic resources. Thus, all modern data transfer tools initiate multiple concurrent socket connections between source and destination for transferring large amounts of files. Numerous studies over the last few decades have attempted several methods to optimize this concurrency value, which balances between high networking resource utilization without creating contention and low system overhead. However, this traditional data transfer architecture creates several major challenges for modern high-performance networks. The available hardware resources at HPC clusters and the networks connecting them vary significantly in I/O, computing, and data transmission speeds. With so many different components involved between source and destination, it is almost impossible for all of them to have similar performance for each thread. For example, to transfer data at 100 Gbps, the read speed at the source, the write speed at the destination, and the network paths connecting them must be capable of that. But to achieve 100 Gbps, the required threads for read, write, and network might be significantly different. This is because the source or destination might have different thread-level I/O speeds due to hardware specifications (SSD or HDD) or resource contention from background I/O, computing, or networking jobs. Similar issues exist for networks too; additionally, system administrators often restrict per-connection speed for fair bandwidth sharing among all applications.

Current data transfer tools use socket connection threads for all read, write, and transfer operations. As a result, the lowest-performing component always determines the required concurrency level for all other components. Thus, if a sysadmin throttles per-connection speed at 1 Gbps on any link or in any intermediate path, we would need 100 parallel socket connections for full utilization, and existing tools will set the read and write concurrency to 100 (where 8–10 would suffice) because the monolithic design couples all components. This not only creates significant overhead on end systems, but unnecessary concurrency massively degrades the performance for all existing processes. The adverse impacts of monolithic design on modern HPC infrastructures have been extensively explored in the Marlin~\cite{marlin} study.

Therefore, rethinking the traditional architecture is necessary for modern HPC infrastructures. Several ongoing projects, FDT~\cite{fast_data_transfer}, mdtmFTP~\cite{The_MDTM_project}, and Marlin~\cite{marlin} are already working on decoupling the read, write, and network components to independently tune them. We refer to this decoupling aspect as modular architecture throughout this paper. As of this writing, neither FDT nor mdtmFTP has publicly available manuscripts or software; therefore, we do not include them in direct comparisons.

% \begin{figure*}[htbp]
%     \centering
%     \subfigure[Marlin]{%
%         \includegraphics[width=0.4\textwidth]{images/marlin_bad.pdf}%
%         \label{fig:marlin_bad}%
%     }
%     \quad
%     \subfigure[Falcon]{%
%         \includegraphics[width=0.4\textwidth]{images/falcon_bad.pdf}%
%         \label{fig:falcon_bad}%
%     }
%     \caption{Marlin's concurrency is unstable due to not capturing the full picture of the memory buffer dynamics, while Falcon also suffers from this phenomenon, resulting in unstable throughput.}
%     \label{fig:bad}
% \end{figure*}

% \begin{figure*}[t!]
%     \centering
%     \begin{subfigure}[t]{0.5\textwidth}
%         \centering
%         \includegraphics[height=1.2in]{marlin_bad.pdf}
%         \caption{Lorem ipsum}
%     \end{subfigure}%
%     ~ 
%     \begin{subfigure}[t]{0.5\textwidth}
%         \centering
%         \includegraphics[height=1.2in]{falcon_bad.pdf}
%         \caption{Lorem ipsum, lorem ipsum,Lorem ipsum, lorem ipsum,Lorem ipsum}
%     \end{subfigure}
%     \caption{Caption place holder}
% \end{figure*}

We consider file transfer to be a three-step process. First, read threads load files from the source file system into the shared memory of the Data Transfer Nodes (DTNs). Second, the files are sent over the network to the shared memory of the destination DTNs. Finally, write threads sync the incoming files to destination file system. In the following discussion, these steps are referred to as the read, network, and write operations, respectively. Marlin first attempted gradient-based joint optimization for all three components, however the optimizer failed to converge to target solutions~\cite{marlin}. So, Marlin run three independent gradient descent optimizer for separately estimating read, write and network concurrency values. While this simplifies the optimization process, the optimizer failed to consider the dependency among read, write and network processes as shown in Figure~\ref{fig:diagram}. Here, $t_r, t_n, $and $t_w$ refer to the throughput of the read, network, and write operations. Throughout this paper the word \emph{buffer} refers to the application‑level staging directory on each DTN
(a tmpfs mount such as \texttt{/dev/shm}) where file chunks rest temporarily before being flushed to the final filesystem. We do not tune TCP send/receive buffers, so AutoMDT remains agnostic to kernel‑level congestion control and can be deployed without any transport layers tuning. As seen in the diagram, each stage's throughputs are not independent. If we do not combine all three stages during optimization, the process can be misguided, and take a long time to converge (Figure~\ref{fig:no_bn}). Alternatively, when all operations are optimized together using multivariate gradient descent, the read throughput will initially increase with increased concurrency because the buffer is not full. Observing this trend, the gradient descent may continue increasing the read concurrency. However, after the buffer becomes full, further increases become unnecessary. Similarly, first increasing the network or writing concurrency does not produce good results because the buffer is empty. In that case, the increase may slow down or even decrease when it should be increased a few steps later. Multivariate gradient descent gets stuck to local optima at the beginning (increase read, while maintaining steady network and write concurrency), and never recovers from that. \emph{That's why joint optimization failed in Marlin. To solve this problem, we next build the simulator to emulate intricate relations among these components and train our agent to learn the overall dynamics first for more effective optimization.}
%\vspace{-2mm}

\section{AutoMDT: Automated Modular Data Transfer Optimization}\label{automdt}
\begin{figure*}[ht]
    \centering
    \includegraphics[width=0.85\textwidth]{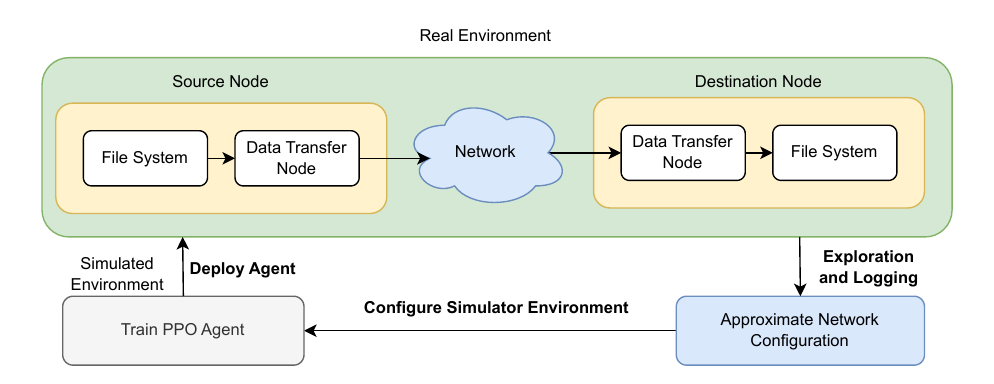}
    \caption{AutoMDT introduces offline training of a deep reinforcement learning agent to quickly learn the behavior and memory buffer dynamics of the real environment.}
    \label{fig:workflow}
\end{figure*}

We formulated the problem as an optimization task involving complex dynamics between several variables. As demonstrated in Figure \ref{fig:workflow}, it relies on training a Proximal Policy Optimization (PPO) agent in a simulated environment rather than through online training, which allows for a substantial reduction in convergence time during training. For example, previous work by Hasibul et al. \cite{drl} applied an online training approach to estimate a single concurrency value without separating network and I/O tasks. Their method required about 28 hours of online training for $5000$ iterations to estimate the optimal value for concurrency. Training for data transfer tasks poses unique challenges because we can only evaluate one network configuration at a time, and we have to wait at least 3 to 5 seconds to get stable metrics for that configuration. As a result, training for data optimization agents takes extremely long times, and we have to repeat this for every network. Now, each additional parameters increases the search space exponentially. As we have three parameters, our models take $15000-30000$ episodes (each with 10 steps) in different testbed settings. That means, if we followed a fully online training approach, it would take approximately five days to train the agent ($150000*3 = 450000sec$). Not only does that create a significant computational burden, it also heavily wastes network bandwidth; in a 100 Gbps network that translates to almost $450000*12.5 \approx5.62PB$ of data transfers.  This motivates us to develop a testbed simulator that replicates the key I/O and network dynamics shown in Figure~\ref{fig:diagram}. This simulator enables us to train the DRL agent offline in a drastically shorter time. The training in simulator works mainly because, unlike previous studies, the DRL-based optimizer no longer directly predicts the concurrency using network metrics, rather tries to learn a generalized dynamics of systems and networks. And, from there trying to make a decision to increase or decrease the concurrency values. Next, we discuss the DRL agent and simulator architectures in detail.

\subsection{Exploration and logging}

We begin with a 10‑minute “random‑threads” run.  
Every second we record the current thread counts  
\(\langle n_r,n_n,n_w\rangle\) and the corresponding per‑stage
throughputs \(\langle T_r,T_n,T_w\rangle\).
From the log we keep
\[
B_r=\max T_r,\quad
B_n=\max T_n,\quad
B_w=\max T_w,
\]
\[
TPT_r=\max \frac{T_r}{n_r},\quad
TPT_n=\max \frac{T_n}{n_n},\quad
TPT_w=\max \frac{T_w}{n_w},
\]
and define the end‑to‑end bottleneck
\(b=\min\{B_r,B_n,B_w\}\). Here $B_i$ represents the bandwidth and $TPT_i$ represents throughput per thread, where $i$ can be r, n, or w.

Assuming near‑linear scaling up to the bottleneck, the thread counts
needed to hit \(b\) are
\[
n_r^{\star}= \frac{b}{TPT_r},\quad
n_n^{\star}= \frac{b}{TPT_n},\quad
n_w^{\star}= \frac{b}{TPT_w}.
\]
We will use these values later during the offline training phase.

\subsection{Utility Function}
Since systems or network conditions change dynamically, we proposed a new utility function to offer a generalized reward that adapts to any environment. This function helps the DRL agent update its model and adjust to new network conditions effectively. The utility function aims to maximize throughput while minimizing the number of parallel streams. Our utility function is defined as:
\[
U(n_i, t_i) = U_{read}(t_r, n_r) + U_{network}(t_n, n_n) + U_{write}(t_w, n_w)
\]

Here, \(t_i\) and \(n_i\) represent the throughputs and concurrency values for the read, network, and write operations, denoted by \(t_r, t_n, t_w\) and \(n_r, n_n, n_w\) respectively. The terms \(U_{read}, U_{network}\), and \(U_{write}\) represent the utility achieved by each operation. They are defined as:
\begin{align*}
U_{read}(t_r, n_r) &= \frac{t_r}{k^{n_r}}, \\
U_{network}(t_n, n_n) &= \frac{t_n}{k^{n_n}}, \\
U_{write}(t_w, n_w) &= \frac{t_w}{k^{n_w}}.
\end{align*}

Higher values of \(t_r, t_n, t_w\) increase the utility, but they often require higher values of \(n_r, n_n, n_w\), which increase the penalty through the term \(k^{n_i}\). In this way, we ensure that there is a global maximum, which becomes the goal for the agent. The value of \(k\) is significant as it balances between resource usage and throughput. A higher value of \(k\) encourages fewer threads, while a lower value of \(k\) focuses on achieving the highest possible throughput, even if more resources are used. This is a tunable parameter to control the aggressiveness of the optimization agent and can be set during runtime. In a simple sweep across several links (1–25 Gbps), the sweet spot was just above 1 (specifically 1.02). We therefore fix \(k=1.02\) for all results in this paper.

\begin{algorithm}\label{algo:simulator}
\caption{I/O and Network Dynamics Simulator}
\begin{algorithmic}[1]
\State \textbf{Initialization:} Set buffer capacities, buffer usages, throughputs per thread for three operations $TPT_i$, bandwidths, initial thread counts, and simulation duration $T_{\text{end}}$.
\medskip
\Function{Task}{$t$, thread\_type}
    \State $throughput\_increase \gets 0$
    \State Task duration, $d_{task} \gets 0$
    \If {thread\_type = "read"}
        \If {sender buffer is not full}
            \State Compute throughput increase.
            \State Compute $d_{task}$ according to $TPT_r$.
            \State Update read throughput and sender buffer usage.
        \EndIf
    \ElsIf {thread\_type = "network"}
        \If {sender buffer $>$ 0 \textbf{and} receiver buffer is not full}
            \State Compute throughput increase.
            \State Compute $d_{task}$ according to $TPT_n$.
            \State Update network throughput, decrease sender buffer, and increase receiver buffer.
        \EndIf
    \ElsIf {thread\_type = "write"}
        \If {receiver buffer $>$ 0}
            \State Compute throughput increase.
            \State Compute $d_{task}$ according to $TPT_w$.
            \State Update write throughput and receiver buffer usage.
        \EndIf
    \EndIf
    \State $t_{\text{next}} \gets t + d_{task} + \epsilon$
    \State \Return $t_{\text{next}}$
\EndFunction
\medskip
\Function{Get\_Utility}{new\_threads}
    \State Reset throughput counters.
    \State Schedule initial tasks for each thread in \texttt{new\_threads} with $t = 0$.
    \While {the task queue is not empty}
        \State Pop $(t, \; thread\_type)$ from the queue.
        \State $t_{\text{next}} \gets$ \Call{Task}{$t$, thread\_type}
        \If {$t_{\text{next}} < T_{\text{end}}$}
            \State Add $(t_{\text{next}}, \; thread\_type)$ to the queue.
        \EndIf
    \EndWhile
    \State Normalize throughputs by their finish times.
    \State Compute reward
    \State Update the internal simulator state.
    \State \Return reward and other necessary information.
\EndFunction
\end{algorithmic}
\end{algorithm}

\subsection{I/O and Network Dynamics Simulator}
We designed a testbed simulator to train the PPO agent offline. The simulator is initialized with the buffer capacities at both ends, throughput per thread, bandwidth, and current concurrency values for read, network, and write operations. We assume that an infinite number of files are available to be chunked as needed. 

% To keep it simple, we avoid considering background traffic in the network.

The simulation runs when the $get\_utility$ function (Algorithm 1) is called and simulates one second of transfer operations. During each simulation interval, the throughput counter starts at zero. To make the code both efficient and practical, we use a priority queue instead of threads. The queue is sorted by time, and when a task (representing a thread's work) is popped from the queue, the simulator checks if transferable data is available. If data is available and the buffers are not full, the thread executes its task. If no data is available or the buffer is full, the task is returned to the queue with a small time increment $\epsilon$ added, so it can retry after a short delay.

Once the queue is empty, we normalize the throughput to determine the exact amount achieved in one second. Finally, we calculate the reward using the utility function. The current state values are saved for future use, and all necessary metrics are returned to the PPO agent.

\subsection{PPO Agent Architecture}
We choose policy-driven DRL as we do not want the agent to overfit to specific actions in the simulator, but to learn generalizable policies that capture different dynamics. PPO~\cite{schulman2017ppo} is the most widely used policy-based DRL, so we choose PPO for our optimizer agent. The PPO agent follows usual PPO architecture and several design choices were made to form the states, actions, and both the policy (actor) and value (critic) networks. The design of the states and actions is key, as one of the main responsibilities of a PPO agent is to learn the mapping from states to actions. This mapping is learned through an actor network, which determines the best actions to take, and a critic network, which estimates the value of each state. In the following sections, we describe these four components in detail.

\subsubsection{State Space}
Defining the state space is one of the most important parts of a PPO design. A proper state space can guide the agent effectively, while including too many states may lead to unnecessary exploration. Our challenge was to design a state space that helps the agent perform well in diverse network scenarios during offline training. For example, if we only consider concurrent thread counts and the corresponding throughput, the agent may get confused because the same state can yield different rewards due to the dynamic nature of the memory buffer discussed in the motivation section.

To address this, we found that the most important information is the available buffer space at both the sender and the receiver ends. Every DTN measures its available buffer space with a system call and the receiver sends the result to its peer over the RPC channel. We designed the state space to include the current thread counts, throughputs, and the amount of unused buffer at both the sender and the receiver. These values give the state a solid foundation and help the model differentiate new scenarios from those it has already seen.

\subsubsection{Action Space}
% The action space can be designed in various ways. One approach is to define the actions as discrete changes in concurrency for each operation. Hasibul et al. \cite{drl} used this technique. Their actions include adding 5 or 1 stream, keeping the concurrency unchanged, or removing 5 or 1 stream. Although this design provides flexibility, the agent takes a long time to learn. This is evident from their results, where it took 28 hours of online training just to learn the dynamics of a single variable, whereas we have 3 variables to consider.

% In our final design, 
We define the concurrency values directly as actions. The policy network has three heads for read, write and network values, each predicting the corresponding thread count. This design allows the agent to directly map a state to an action, which helps the model learn and converge faster without requiring a large amount of information.

\subsubsection{Policy Network}
The policy network is designed to predict actions directly from the current state using a series of fully connected layers enhanced by residual connections. Initially, the input is embedded into a 256-dimensional space using a linear layer followed by a \texttt{tanh} activation. The embedded representation then passes through a sequence of three residual blocks. Each residual block comprises two linear transformations interleaved with layer normalization and ReLU activations, along with a skip connection that adds the input directly to the output. This architecture facilitates better gradient flow and allows the network to learn complex state representations efficiently. The output of the residual blocks is processed by a \texttt{tanh} function before being fed into a linear layer to compute the mean of the action distribution. Simultaneously, we clamp the trainable log–standard-deviation parameter to a reasonable range and exponentiate it to produce the standard deviation. Together, these outputs allow the model to sample actions from a normal distribution, effectively capturing both the deterministic mapping and the inherent uncertainty in the environment.

\subsubsection{Value Network}
The value network is responsible for estimating the expected return for a given state, a critical component for calculating advantages in the PPO framework. In this design, the state is first transformed into a 256-dimensional feature space via a linear layer, followed by a \texttt{tanh} activation. To further refine this representation, the network employs two residual blocks, each built using a custom residual block structure with Tanh activations. These residual blocks consist of two sequential linear layers and incorporate a skip connection to enhance feature propagation and mitigate vanishing gradients. Finally, the refined features are passed through a linear layer to produce a single scalar value as the estimated return. This residual-based architecture improves the stability and accuracy of the value estimates, particularly in complex and dynamic environments.

\begin{algorithm}
\caption{PPO training for optimizing thread allocation}
\label{algo:ppo}
 % with convergence criterion
\begin{algorithmic}[1]
\Require Optimization environment $\mathcal{E}$, maximum step per episode $M$, maximum episodes $N$, learning rate $\alpha$, discount factor $\gamma$, clipping threshold $\epsilon$, theoretical maximum reward $R_{max}$
\Ensure Save the best policy $\pi_\theta(s)$ and value network $V_\phi(s)$ that optimize thread allocation
\State Initialize parameters $\theta$ for policy and $\phi$ for value network
\State Initialize memory $\mathcal{M}$
\State Set episode counter $n \gets 0$, best reward $R^* \gets 0$, stagnant counter $c \gets 0$
\While{$n < N$}
    \State Reset environment: $s \gets \mathcal{E}.\text{reset()}$
    \State Set step counter $m \gets 0$, episode reward $r_{ep} \gets 0$, and clear memory $\mathcal{M}$
    \While{$m < M$}
        \State Compute mean and standard deviation: $(mean,\, std) \gets \pi_\theta(s)$
        \State Sample action: $a \sim \mathcal{N}(mean, std)$
        \State Execute action: $(s', r, done) \gets \mathcal{E}.\text{step}(a)$
        \State Store $(s, a, r)$ in $\mathcal{M}$
        \State Update state: $s \gets s'$
        \State Update episode reward: $r_{ep} \gets r_{ep} + r$
        \State Increment $m \gets m+1$
    \EndWhile
    \State Let $states, actions, rewards \gets \mathcal{M}$
    \State Compute discounted returns $G_t = r_t + \gamma\,G_{t+1}$, $\forall t$
    \State Compute $(mean,\, std) \gets \pi_\theta(states)$
    \State Define distribution $\mathcal{D} \sim \mathcal{N}(mean,\, std)$
    \State Compute $entropy \gets$ sum of entropies of $\mathcal{D}$ over action dimensions
    \State Compute policy ratio:
            \[
            r_t = \frac{\pi_\theta(a_t | s_t)}{\pi_{\theta_{\text{old}}}(a_t | s_t)}
            \]
    \State Compute advantages: $A_t = G_t - V_\phi(s_t)$
    \State Compute surrogate terms:
    \[
    surr1 \gets r_t\cdot A_t, \quad surr2 \gets \text{clip}(r_t,\, 1-\epsilon,\, 1+\epsilon)\cdot A_t
    \]
    \State Actor loss: $\mathcal{L}_{actor} \gets -\min(surr1, surr2)$
    \State Critic loss: $\mathcal{L}_{critic} \gets 0.5\,\text{MSE}(G_t,\, V_\phi(s_t))$
    \State Total loss: $\mathcal{L} \gets \mathcal{L}_{actor} + \mathcal{L}_{critic} - 0.1\,entropy$
    \State Backpropagate $\mathcal{L}$ and update parameters using Adam optimizer
    \State Update old policy: $\pi_{\theta_{\text{old}}} \gets \pi_\theta$
    \If{$r_{ep} > R^*$}
        \State $R^* \gets r_{ep}$, $c \gets 0$, Save model
    \Else
        \State $c \gets c+1$
    \EndIf
    \If{$R^* \ge 0.9\,R_{max} \And c \ge 1000$}
        % \If{$c \ge 1000$}
            \State \textbf{break} \Comment{Convergence achieved}
        % \EndIf
    \EndIf
    \State Increment episode counter: $n \gets n+1$
\EndWhile
\end{algorithmic}
\end{algorithm}

\subsection{Training Algorithm}
The training algorithm is responsible for updating the policy and value networks to optimize concurrency allocation. It takes as input the optimization environment \(\mathcal{E}\), maximum step per episode \(M\), maximum episodes \(N\), learning rate \(\alpha\), discount factor \(\gamma\), clipping threshold \(\epsilon\), and theoretical maximum reward $R_{max}$. The training process begins by initializing the parameters of the policy and value networks. The algorithm then runs for \(N\) episodes unless the convergence criterion is met. After each episode, the optimization environment is reset to test the networks with a new state consisting of a new set of randomly initialized threads, and both the step counter and memory are reset.

For each episode, the algorithm runs a loop for \(M\) steps. In each step, the agent selects an action using the policy network, explores the environment, collects observations, and stores them in memory. Once the exploration phase is complete, the algorithm computes the discounted returns, the advantages, and the entropy of the action distribution. 

The overall loss function for the policy network combines three components. First, the actor loss guides the policy update by comparing the probabilities of actions under the new and old policies, using a clipping mechanism to limit large updates and ensure stability. Second, the critic loss minimizes the difference between the predicted state values and the computed discounted returns, which helps the value network to estimate state quality accurately. Finally, an entropy regularization term is incorporated to encourage exploration by preventing the action distribution from becoming overly deterministic. After computing these losses, the total loss is back-propagated, and the network parameters are updated using the Adam optimizer.

The convergence criterion is straightforward. We first calculate a theoretical maximum achievable reward.
With the thread counts we obtained from the logging and exploration phase and the penalty factor \(k=1.02\), the highest reward achievable is
\[
R_{\max}=b\bigl(k^{-n_r^{\star}} + k^{-n_n^{\star}} + k^{-n_w^{\star}}\bigr).
\]
% Assuming every stage runs with its optimal thread count and overall throughput is capped by the slowest stage, the reward is bounded by
% \[
% R_{\max} \;=\;
% \sum_{i \in \{r,n,w\}}
% \frac{\min\{B_r,\,B_n,\,B_w\}}{k\,n_i^{optimal}}.
% \]

% Here \(B_r, B_n, B_w\) are the bandwidths for read, network, and write, and \(k>1\) is the concurrency‑penalty factor.
If the agent reaches 90\% of $R_{\max}$, we consider it to have converged. To allow further refinement, we then wait for an additional 1000 episodes without any improvement in reward after this convergence point before finalizing the model. The training process continues until the convergence criterion is met or the \(N\) episodes are completed, at which point the agent is fully trained and able to efficiently optimize the concurrency allocation.

\subsection{Thread Updates (Production Phase)}
During a production transfer, we load the best checkpoint obtained during offline (simulator) PPO training and re–enter the interaction loop, now with \emph{no preset episode limit} (effectively $N=\infty$) until the current dataset has been fully transferred. In Line 8 of Algorithm~\ref{algo:ppo}, the policy network produces the mean vector $\langle \mu_r, \mu_n, \mu_w \rangle$ and the corresponding log–standard deviations $\langle \sigma_r, \sigma_n, \sigma_w \rangle$. We sample a continuous action from the diagonal Gaussian,
\[
\tilde{a} = \mathcal{N}(\mu, \sigma),
\]
round it to integers to obtain the concurrency tuple
\[
\langle n_r, n_n, n_w \rangle = \mathrm{round}(\tilde{a}),
\]
clamp each component to $[1, n^{\max}]$, and pass this tuple to \textsc{GetUtility}. Instead of sending the values to the simulator, the system performs the data transfer with the updated concurrency settings, probes the achieved throughput, and thereby obtains both the reward and the new states, continuing the loop. Hence, every PPO step explicitly reassigns the concurrency tuple $(n_r,\,n_n,\,n_w)$.
%\vspace{-2mm}

\section{Evaluation}\label{eval}
\begin{figure*}[t]
  \centering

  \begin{minipage}[b]{0.32\textwidth}
    \centering
    \includegraphics[width=\linewidth]{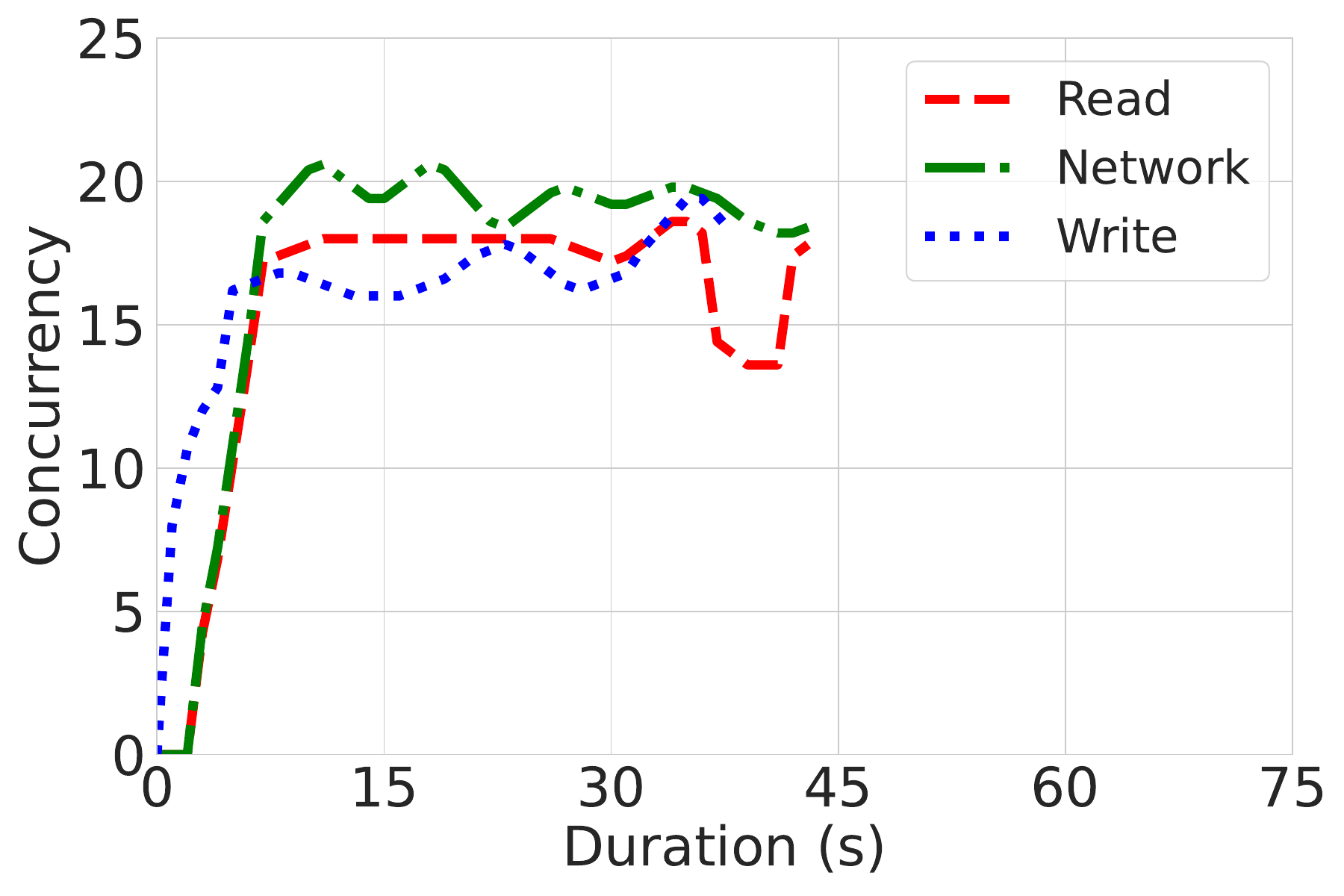}
    \\[-0.3em]
    {\small (a) \name~Concurrency}
  \end{minipage}\hfill
  \begin{minipage}[b]{0.32\textwidth}
    \centering
    \includegraphics[width=\linewidth]{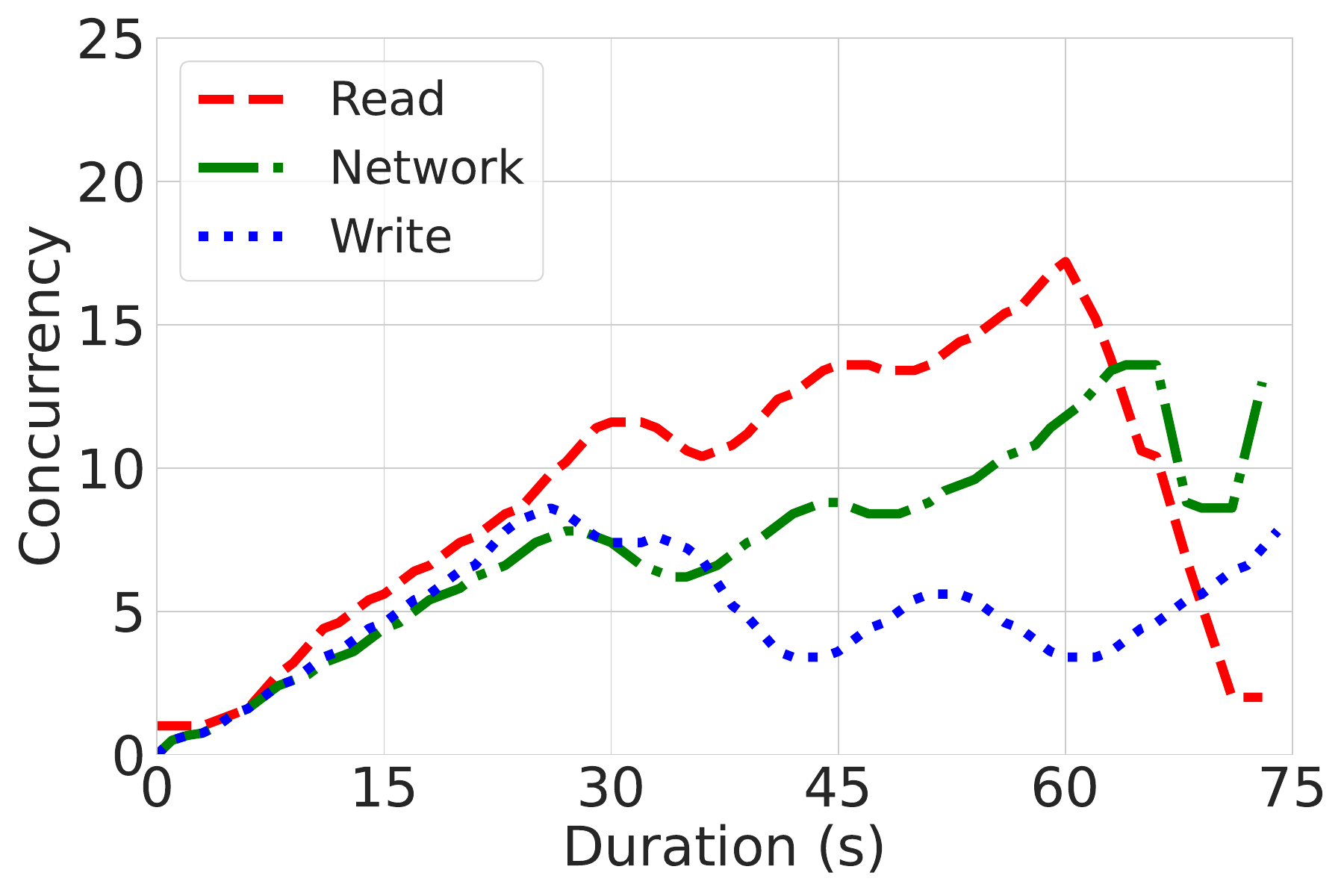}
    \\[-0.3em]
    {\small (b) Marlin Concurrency}
  \end{minipage}\hfill
  \begin{minipage}[b]{0.32\textwidth}
    \centering
    \includegraphics[width=\linewidth]{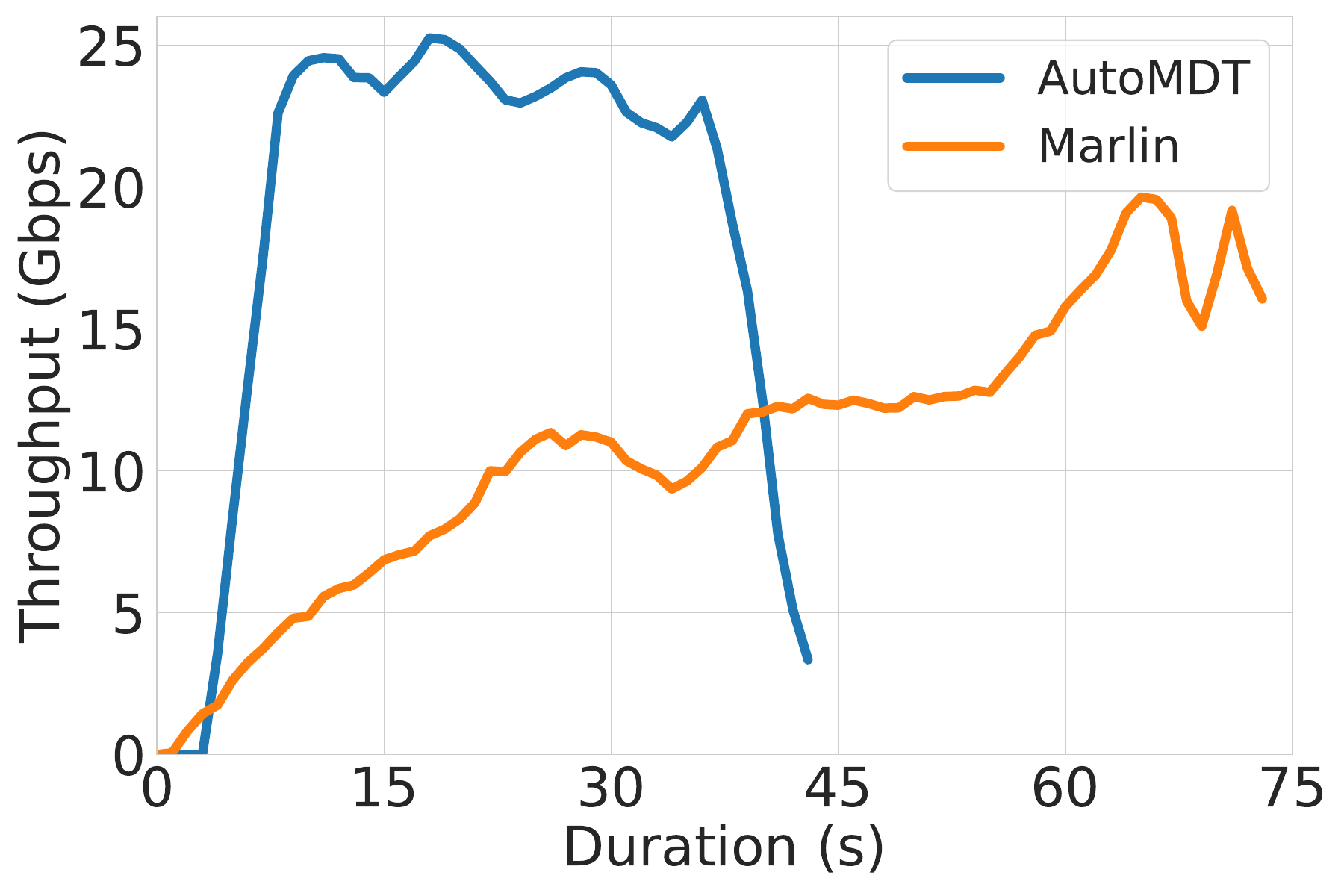}
    \\[-0.3em]
    {\small (c) Throughput Comparison}
  \end{minipage}

  \caption{Performance comparison of \name~and Marlin in Fabric-testbed.
           Marlin takes $\sim$1.7× longer time than \name~to finish the transfer.}
  \label{fig:no_bn}
\end{figure*}

% \begin{table}[ht]
%     \centering
%     \begin{tabular}{llll}
%         \toprule
%         \textbf{Source} & \textbf{Destination} & \textbf{Storage} & \textbf{Bandwidth}   \\
%         \midrule
%         Cloudlab (Wisc) & Cloudlab (Wisc) & RAID-0 SSD & 1.2 Gbps  \\
%         Fabric (BRIST) & Fabric (INDI) & NVMe SSD & 10 Gbps  \\
%         Fabric (NCSA) & Fabric (TACC) & NVMe SSD & 100 Gbps  \\
%         \bottomrule
%     \end{tabular}
%     \caption{Test Environment Specification}
%     \label{tab:setup}
% \end{table}

We demonstrate \name's performance in terms of throughput maximization, concurrency minimization, and stability of the optimization process. We compared the performance of our proposed solution with Marlin~\cite{marlin} and Globus~\cite{globusonline}, a widely used monolithic solution. As mentioned in Section~\ref{motivation}, we were unable to compare with FDT or MDTM as we could not find any of the software available. 

% Notably, Marlin has shown a 2x speedup over the monolithic state-of-the-art model, Falcon \cite{concurrency}. Therefore, we believe that a comparison with Marlin is sufficient to prove the efficiency of AutoMDT.
% testbeds described in Table \ref{tab:setup}

\name~was evaluated on two NSF funded testbeds, CloudLab~\cite{cloudlab} and Fabric~\cite{fabric-2019}. In CloudLab (CloudLab-Wisconsin), both the sender and the receiver use a c240g5 server with an Intel Xeon Silver 4114, 8~GiB RAM, and a 1~Gbps NIC. On the other hand, we used nodes from two different sites, BRIST and INDI, in Fabric \cite{fabric-2019}. Both nodes were configured with 8 cores, 64GB RAM, Dell Express Flash P4510 1TB SFF, and an NVIDIA Mellanox ConnectX-5 NIC. Additionally, we used another pair of fabric nodes from NCSA and TACC. In this setup, we used an NVIDIA Mellanox ConnectX-6 NIC to achieve high network bandwidth. We conducted two types of transfer experiments. The first type of experiments was focused on large files, which were conducted using 1000$\times$1GB randomly generated files. The other types of experiments were focused on mixed datasets to emulate more practical workloads, we used a total of 1TB data consisting of files sizes from 100 KB to 2 GB. 

% The interval for collecting throughput was set to 3 seconds.
% so each iterations took around 4 seconds on average.

\begin{figure}[ht]
    \centering
    \includegraphics[width=0.45\textwidth]{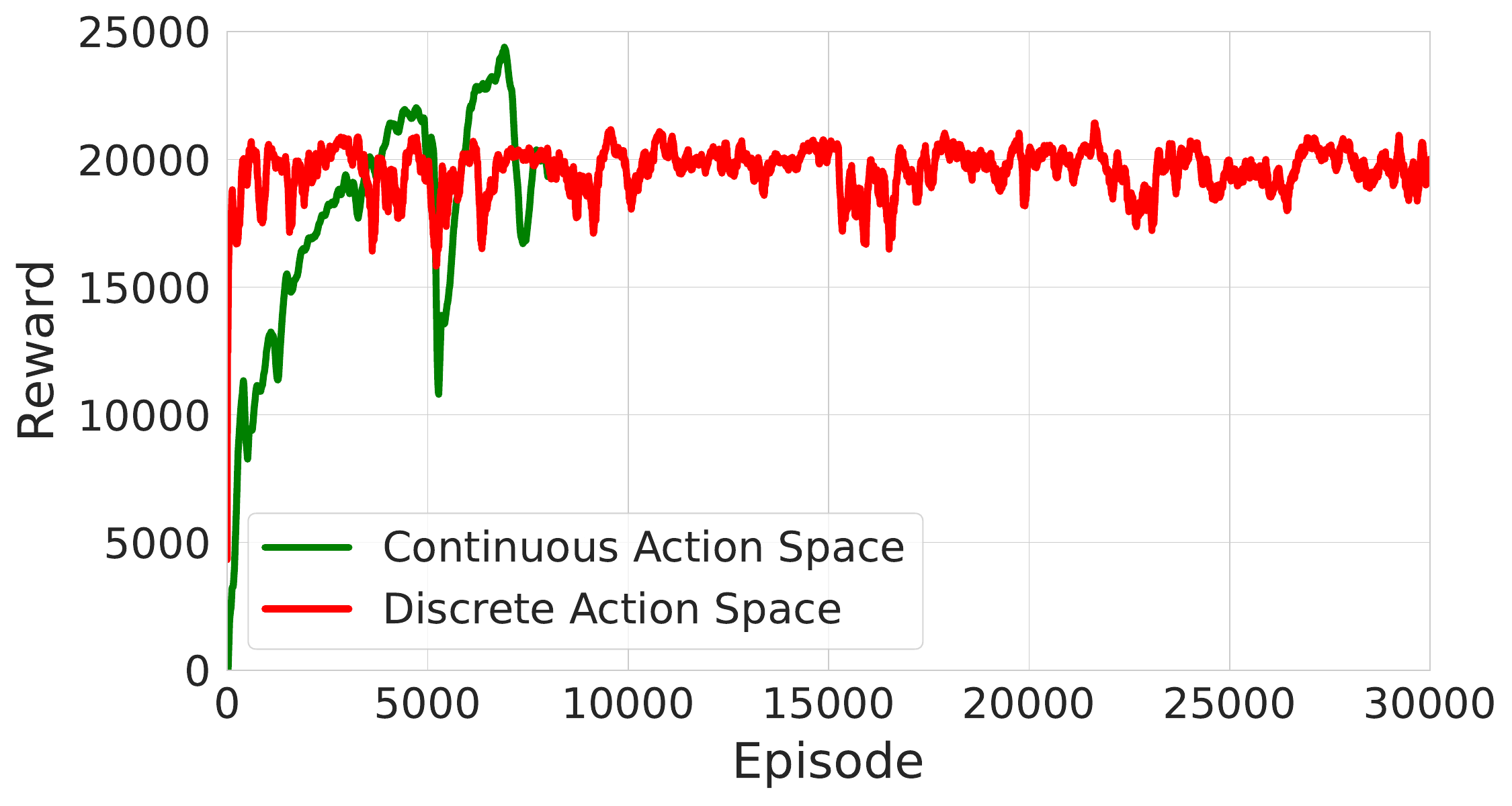}
    \caption{PPO agent with discrete action space failed to achieve convergence.}
    \label{fig:continuous}
\end{figure}

\subsection{Offline Training}
First, we evaluated the offline training with different scenarios and action spaces in both Cloudlab and Fabric testbeds, where we know the expected solutions to test the optimizer efficiency. The average time taken for offline training was around 45 minutes; however, in one experiment, it took as much as 60 minutes to converge. For the experiment illustrated in Figure~\ref{fig:continuous}, we set the maximum number of episodes to 30000 and applied the early stopping condition described in Section~\ref{automdt}. In our experiments, it appeared that reaching the convergence point required approximately \texttt{20150} episodes. Without the simulator, achieving convergence would have taken \texttt{7} days of online training. Please note, each episode contains ten iterations, and each iteration would take 3 seconds in online training.

We also experimented with a discrete action space, as our target concurrency values are discrete. However, the discrete action space failed miserably. Based on Hasibul et al.'s work~\cite{drl}, a significantly more complex state space would be needed to effectively utilize a discrete action space, which in turn would require more complex value and policy models and a longer training time. We settled with continuous spaces, and used rounding to convert the predicted values to integers.

\subsection{Experimental Results}
We ran data transfer from NCSA to TACC using the Fabric testbed. In the first set of experiments, we transferred smaller data size consisting of $100\times1GB$ files. In this experiment, \name~significantly outperformed Marlin. As illustrated in Figure~\ref{fig:no_bn}, Marlin completes the transfer in 74 seconds, whereas \name~takes only 44 seconds (68\% faster). In terms of stability, \name~reached the required concurrency level of 20 in just 7 seconds, Marlin never reached that level. Marlin required 62 seconds to reach 14 (8x slower than \name). Thus, \name~demonstrates superior performance in both speed and stability. \\
% In the 1 TB dataset scenario presented in Table~\ref{tab:thrpt}, AutoMDT finishes more than 100 seconds earlier than Marlin. Although AutoMDT reached the maximum available speed (around 28Gbps) within the first 10 seconds, it took Marlin 215 seconds. 

% To showcase the performance difference with the fixed concurrency options, we conducted experiments with a fixed concurrency of 5. Though it did a better result (420 seconds) than Marlin (442 seconds), it was not able to beat AutoMDT (333 seconds).

\subsubsection{Bottleneck Scenarios}

% \begin{figure*}[ht]
%     \centering
%     \includegraphics[width=0.9\textwidth]{images/combined_9subplots.pdf}
%     \caption{Comparisons of AutoMDT and Marlin. Due to optimizing all the parameters together and leveraging the memory buffer dynamics, AutoMDT quickly identifies the bottleneck operation while maintaining low concurrency for other components. It reaches the optimal solution faster, resulting in improved throughput and better resource utilization compared to Marlin.
% }
%     \label{fig:comparison}
% \end{figure*}

\begin{figure*}[ht]
  \centering

  % Column Titles
  \begin{minipage}[b]{0.32\textwidth}
    \centering
    \textbf{Read I/O Bottleneck}
  \end{minipage}\hfill
  \begin{minipage}[b]{0.32\textwidth}
    \centering
    \textbf{Network Bottleneck}
  \end{minipage}\hfill
  \begin{minipage}[b]{0.32\textwidth}
    \centering
    \textbf{Write I/O Bottleneck}
  \end{minipage}

  \vspace{1em} % Space between column titles and first row

  % Row 1: AutoMDT concurrency
  % \begin{minipage}[b]{0.05\textwidth} % For Y-axis label
  %   \centering
  %   \rotatebox{90}{\small Concurrency (AutoMDT)}
  % \end{minipage}\hfill
  \begin{minipage}[b]{0.32\textwidth}
    \centering
    \includegraphics[width=\linewidth]{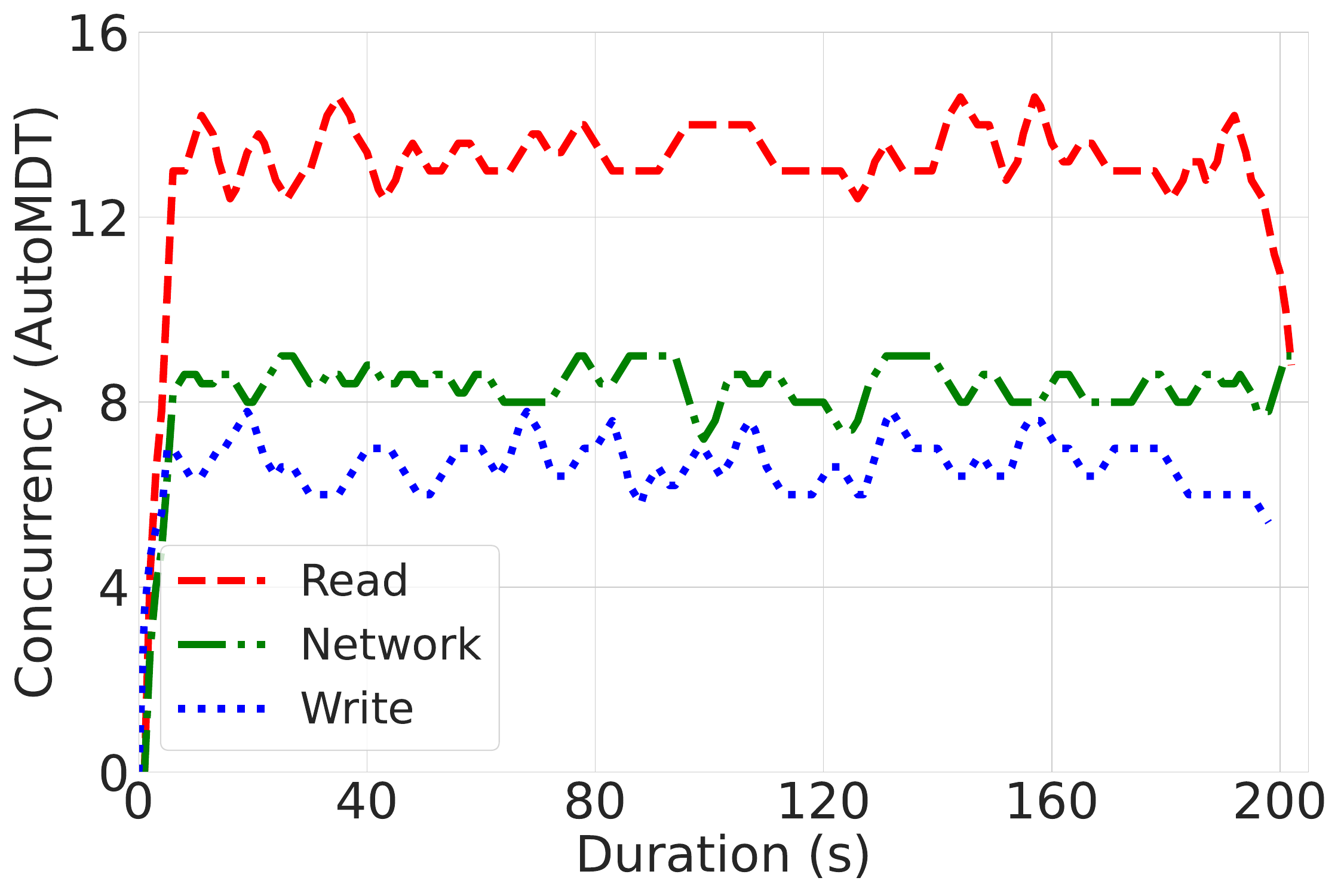}
  \end{minipage}\hfill
  \begin{minipage}[b]{0.32\textwidth}
    \centering
    \includegraphics[width=\linewidth]{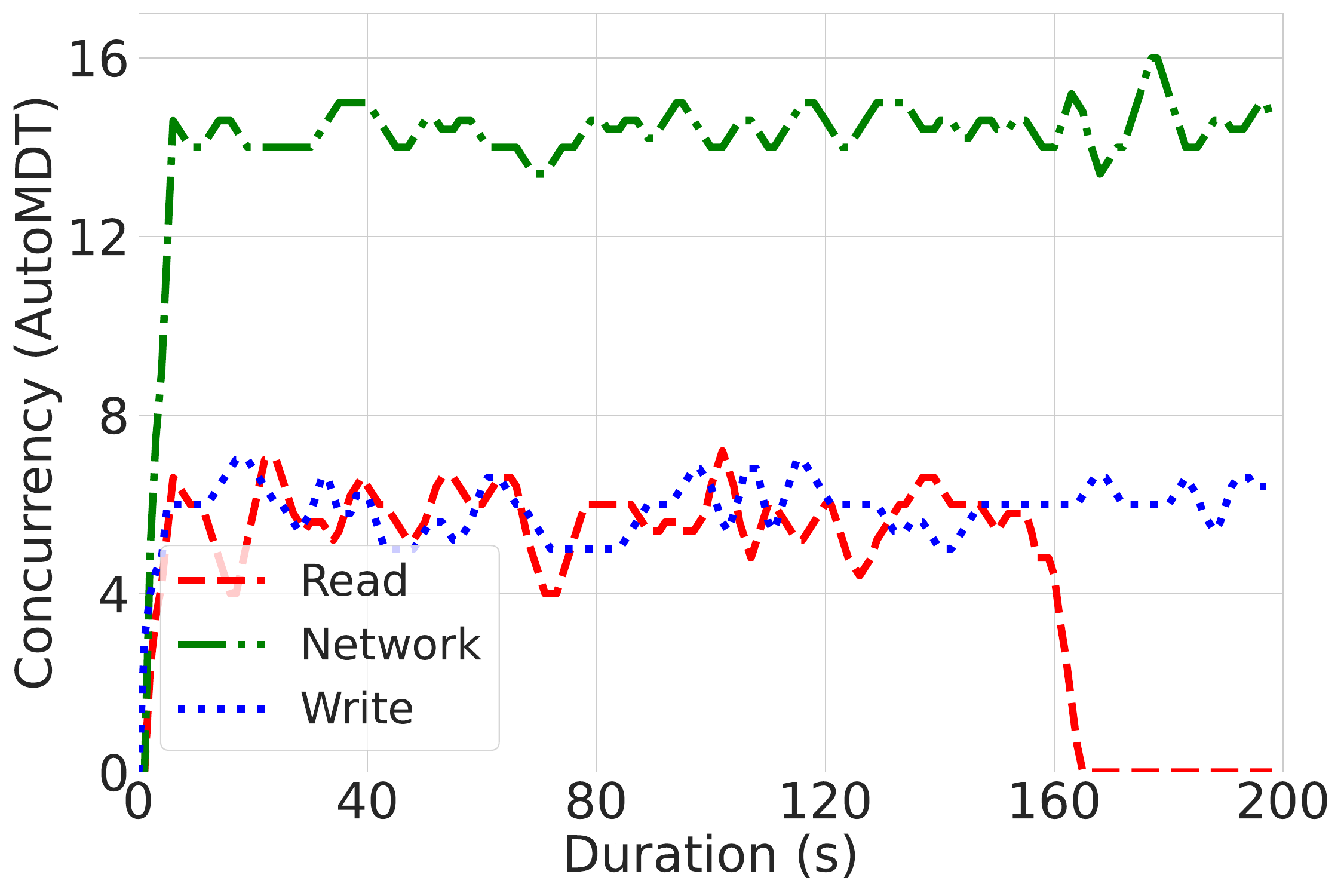}
  \end{minipage}\hfill
  \begin{minipage}[b]{0.32\textwidth}
    \centering
    \includegraphics[width=\linewidth]{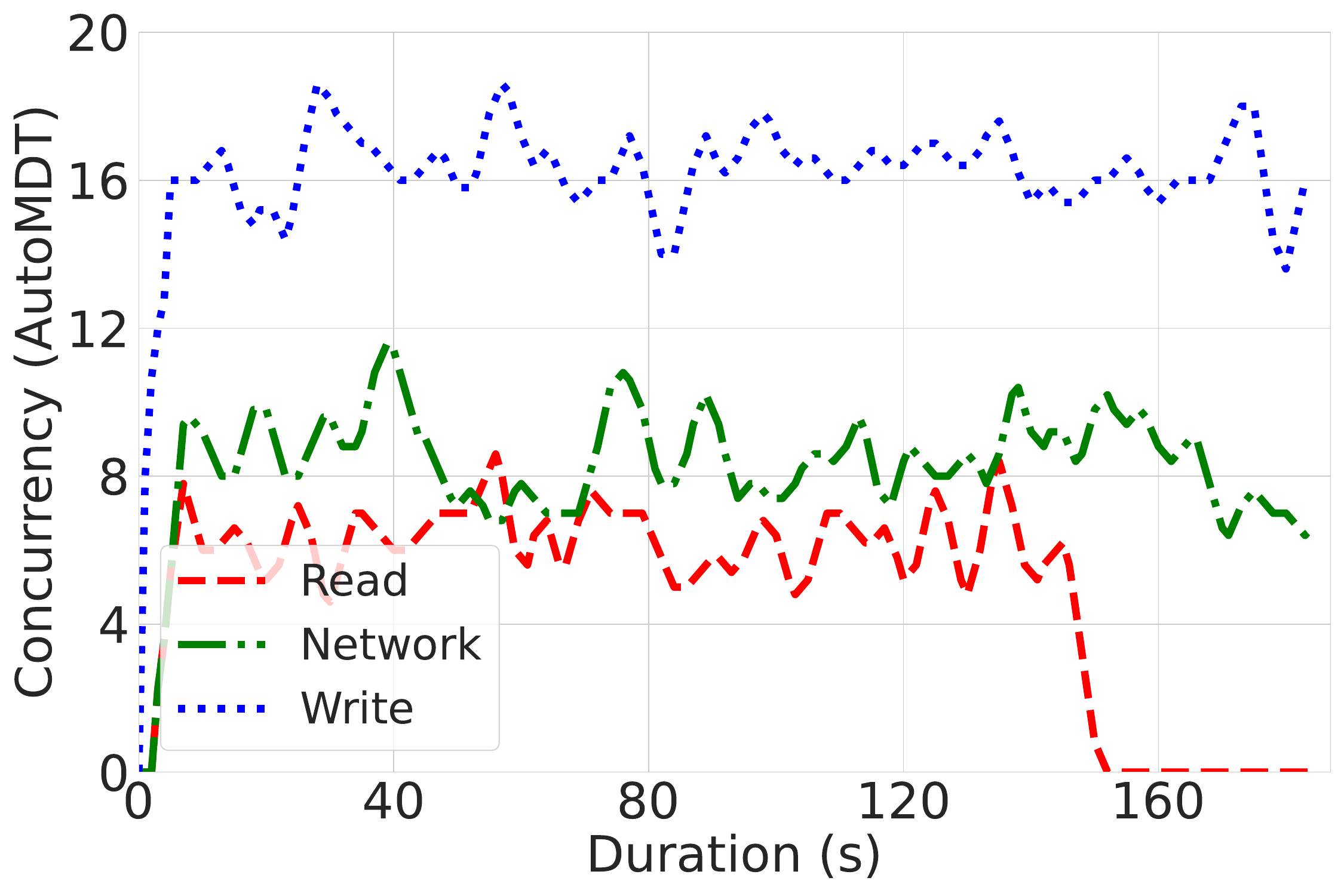}
  \end{minipage}

  \vspace{1em}

  % Row 2: Marlin concurrency
  % \begin{minipage}[b]{0.05\textwidth} % For Y-axis label
  %   \centering
  %   \rotatebox{90}{\small Concurrency (Marlin)}
  % \end{minipage}\hfill
  \begin{minipage}[b]{0.32\textwidth}
    \centering
    \includegraphics[width=\linewidth]{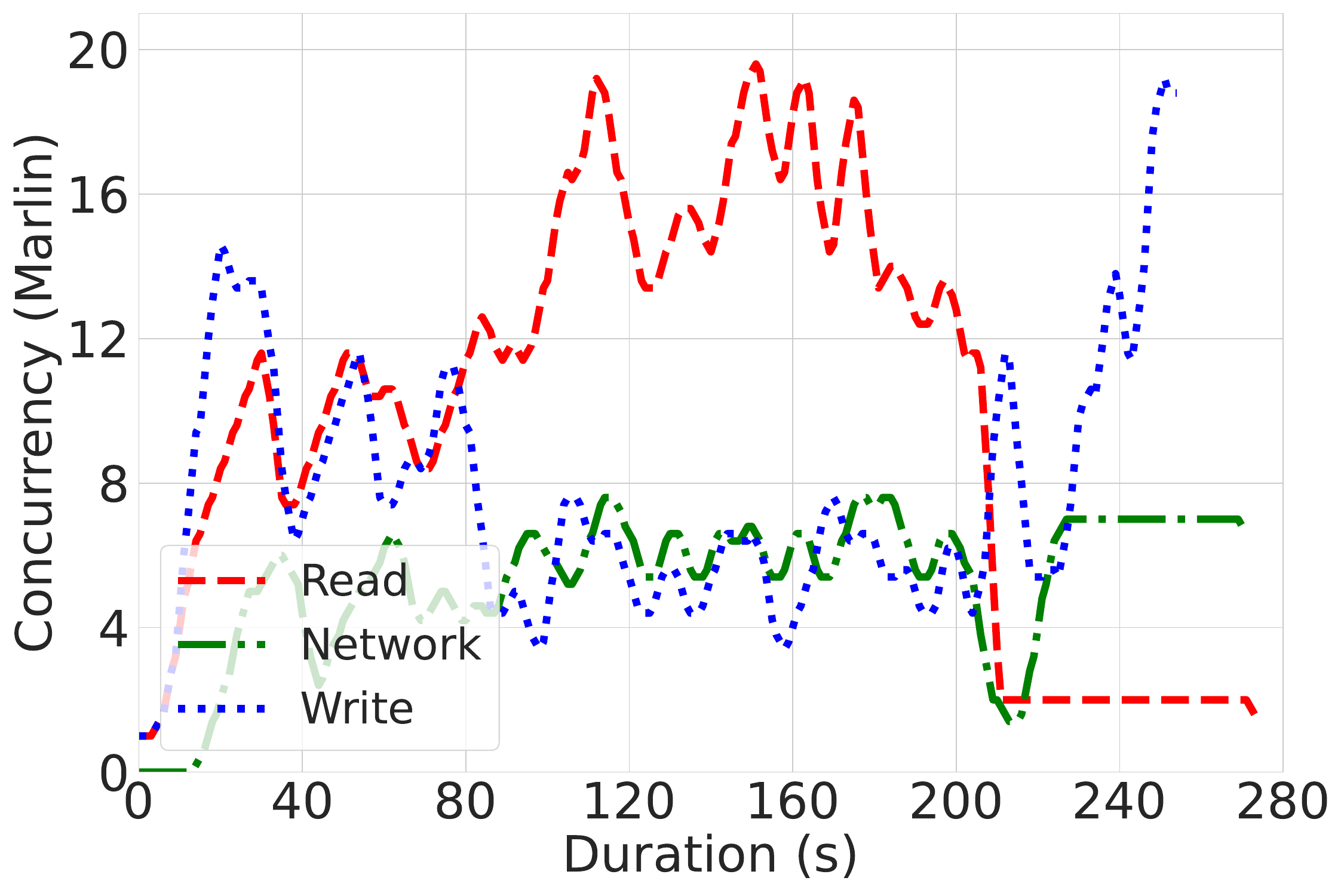}
  \end{minipage}\hfill
  \begin{minipage}[b]{0.32\textwidth}
    \centering
    \includegraphics[width=\linewidth]{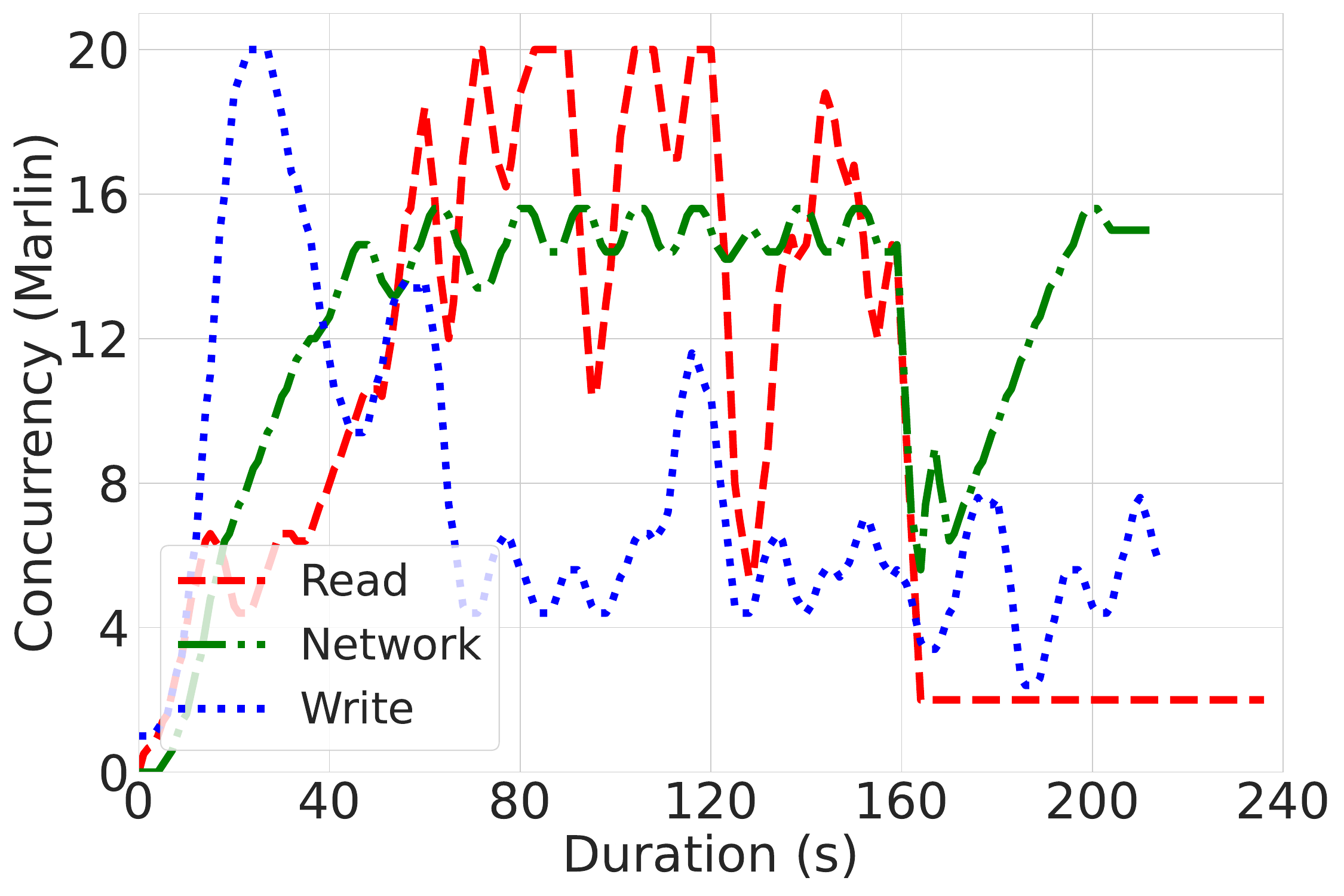}
  \end{minipage}\hfill
  \begin{minipage}[b]{0.32\textwidth}
    \centering
    \includegraphics[width=\linewidth]{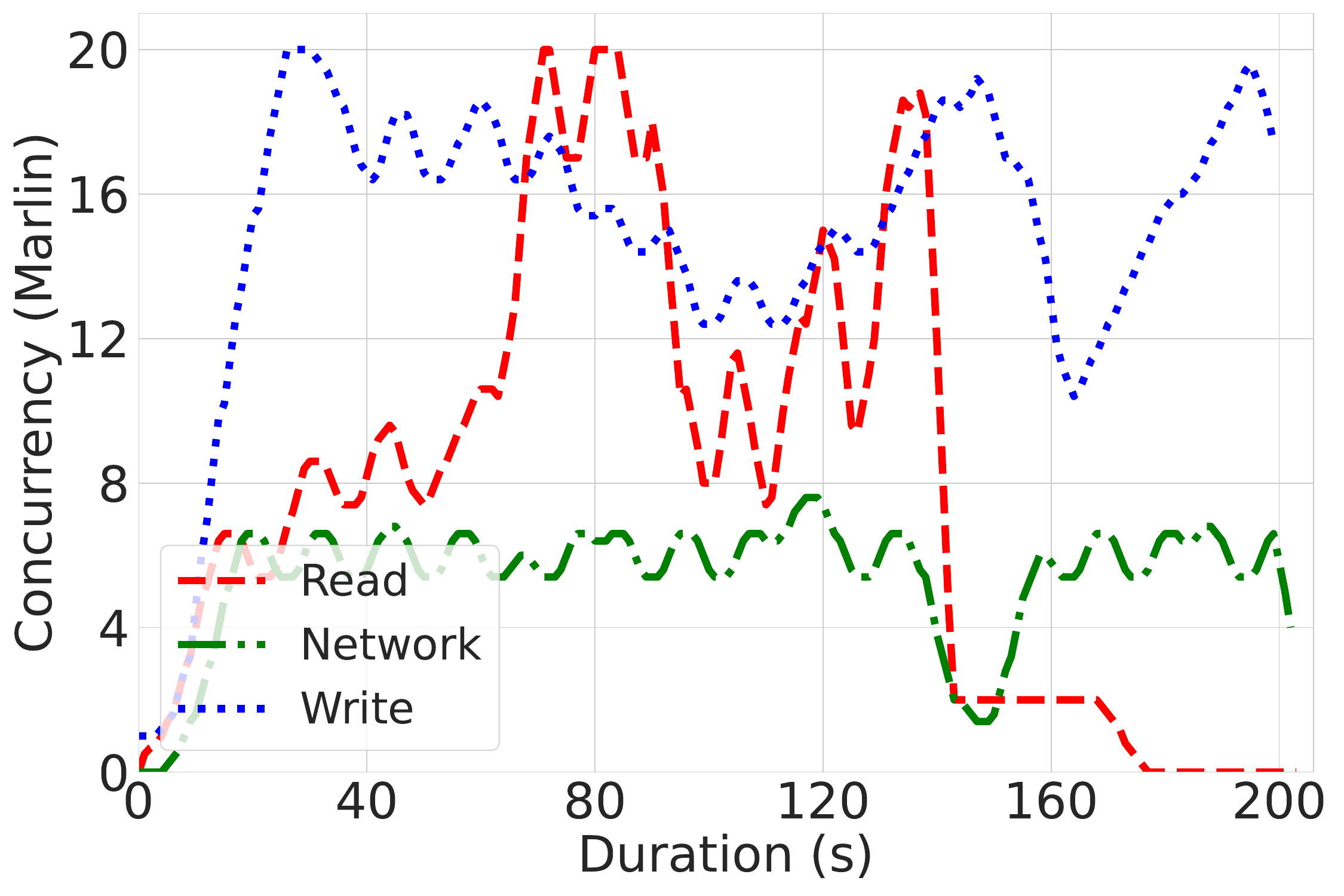}
  \end{minipage}

  \vspace{1em}

  % Row 3: Throughput comparison
  % \begin{minipage}[b]{0.05\textwidth} % For Y-axis label
  %   \centering
  %   \rotatebox{90}{\small Throughput (Mbps)}
  % \end{minipage}\hfill
  \begin{minipage}[b]{0.32\textwidth}
    \centering
    \includegraphics[width=\linewidth]{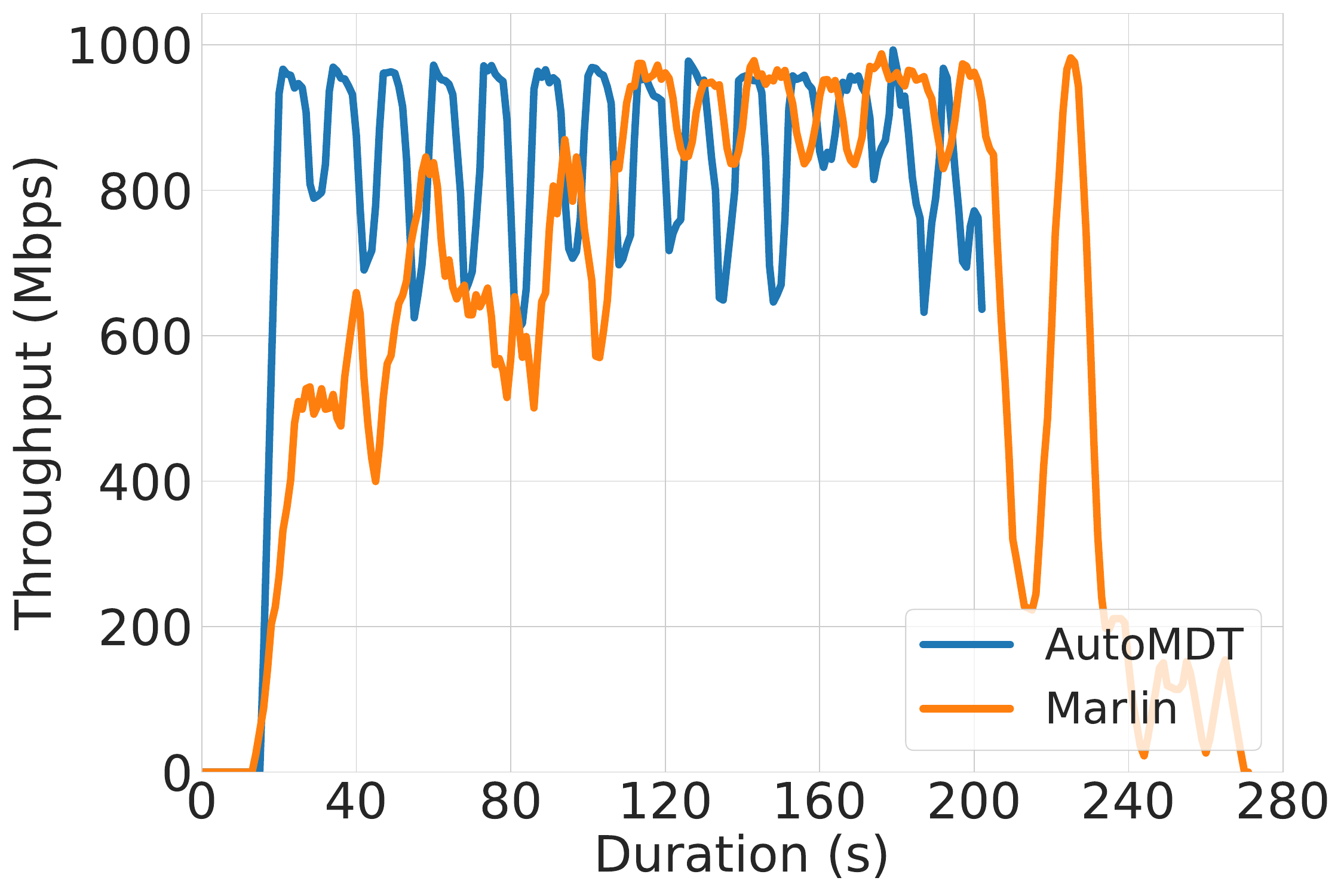}
  \end{minipage}\hfill
  \begin{minipage}[b]{0.32\textwidth}
    \centering
    \includegraphics[width=\linewidth]{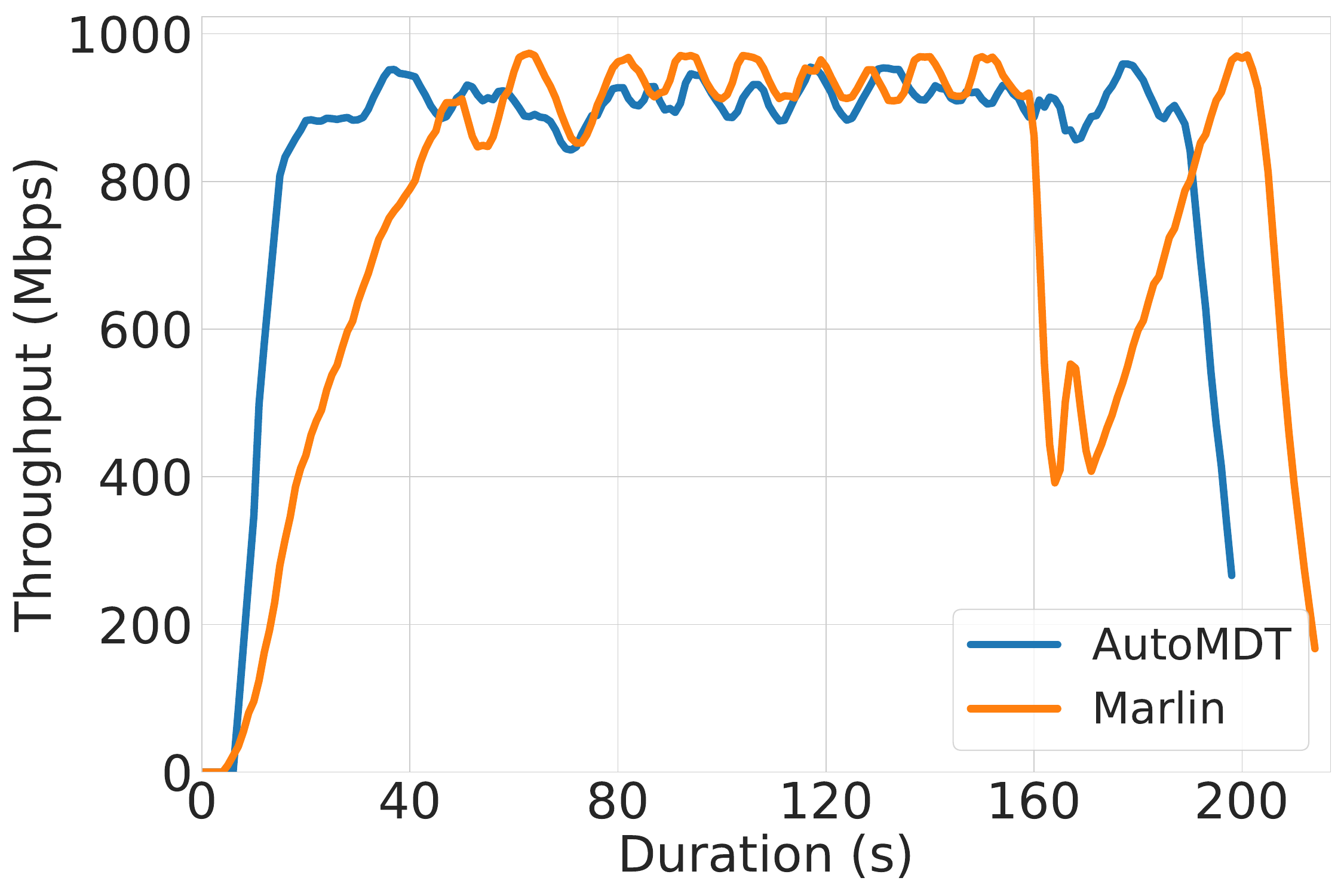}
  \end{minipage}\hfill
  \begin{minipage}[b]{0.32\textwidth}
    \centering
    \includegraphics[width=\linewidth]{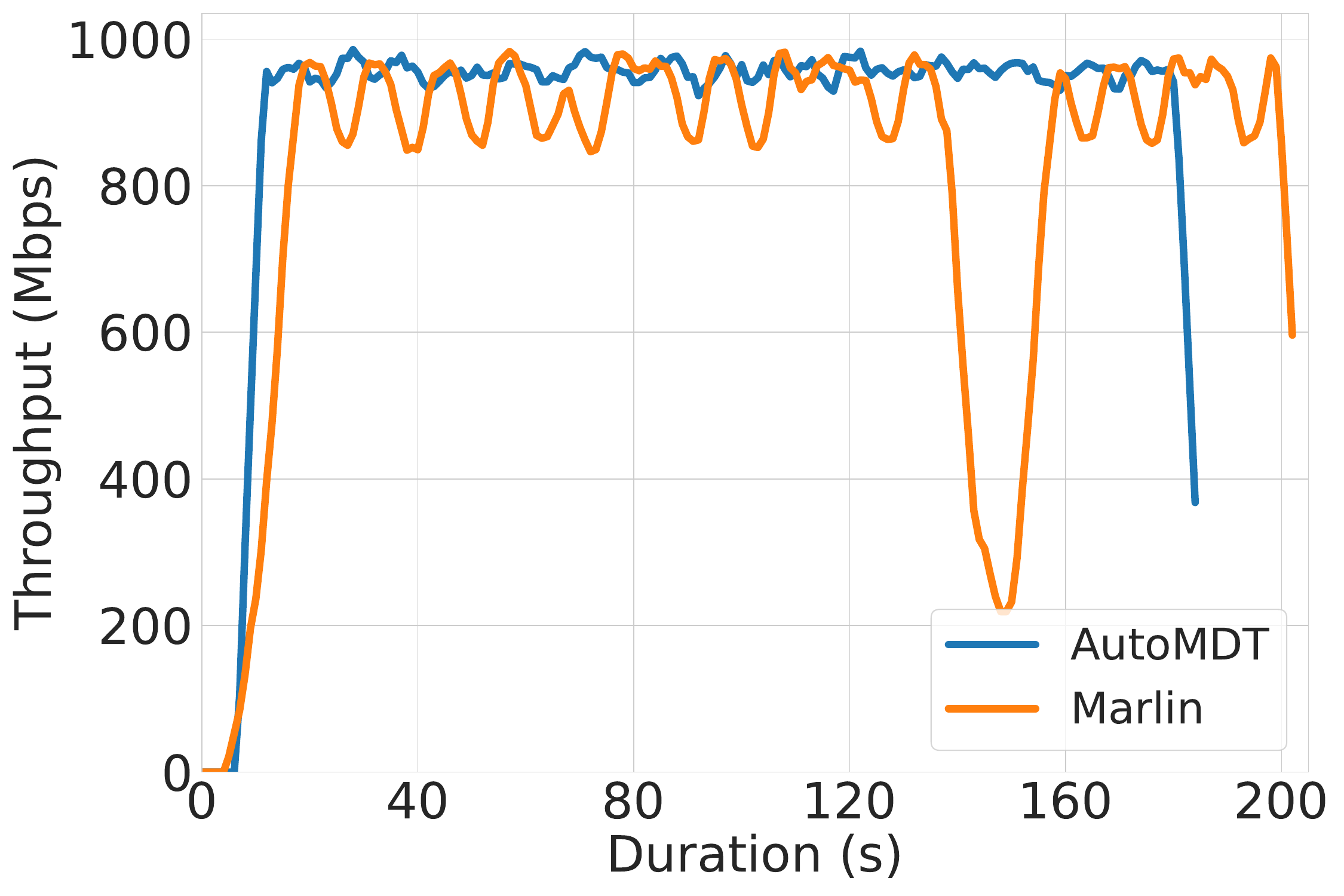}
  \end{minipage}

  \caption{Performance comparisons of AutoMDT (first row) and Marlin (second row). AutoMDT leverages joint-optimization and the memory buffer dynamics to quickly identify the bottleneck component, then increase concurrency accordingly while maintaining low value for other components. It reaches the optimal solution faster, resulting in improved throughput (third row) and better resource utilization compared to Marlin.}
  \label{fig:comparison}
  \vspace{-3mm}
\end{figure*}

To demonstrate the effectiveness of modular architecture, we created bottleneck scenarios on several Fabric node pairs. We manually restricted the throughputs for read, write, and network operations per TCP stream to generate the bottleneck scenarios presented in Figure \ref{fig:comparison}. As discussed in Section~\ref{motivation}, in these bottleneck scenarios, monolithic design optimizers always choose the maximum concurrency required by any of the components to achieve the highest possible utilization. 

To demonstrate the Read bottleneck scenario, we throttled the read threads to 80 Mbps, while write and network connections were limited to 200 Mbps and 160 Mbps, respectively. Given a 1 Gbps network bandwidth, the optimal TCP stream levels for read, network, and write operations are 13, 7, and 5. As shown in the first column of Figure \ref{fig:comparison}, AutoMDT (first row) reaches 13 TCP streams within 6 seconds, whereas Marlin (second row) takes 29 seconds to reach 12 streams. It is also evident that AutoMDT can identify the bottleneck from the beginning, while Marlin's values continue to fluctuate. Consequently, as AutoMDT achieves the optimal numbers earlier, it finishes 68 seconds sooner than Marlin.

Again, we throttled read, network, and write connections to 205 Mbps, 75 Mbps, and 195 Mbps, respectively, to simulate the network bottleneck scenario in the second column of Figure \ref{fig:comparison}. In this case, the optimal TCP stream levels are 5, 14, and 5 for the read, network, and write operations, respectively. The plot shows that AutoMDT achieves stable performance due to its awareness of the memory buffer dynamics, while the read and write concurrency in Marlin remains unstable. Here, AutoMDT reaches 15 in its 3rd second, whereas Marlin reaches 14 in the 42nd second. Consequently, AutoMDT finishes 15 seconds earlier.

In the final scenario, read, network, and write connections were set to 200 Mbps, 150 Mbps, and 70 Mbps, yielding optimal stream counts of 5, 7, and 15 for the respective operations, as shown in the third column of Figure \ref{fig:comparison}. Again, with very stable optimization, AutoMDT finishes 17 seconds earlier than Marlin.

% \begin{figure*}
%   \centering
%   % Row 1: AutoMDT and Marlin Concurrency Plots side by side
%   \begin{minipage}[b]{0.4\textwidth}
%     \centering
%     \includegraphics[width=\textwidth]{images/concurrency_ft_vs_wo_AutoMDT_Offline_network_bn.pdf}
%     \vspace{0.5em}
%     \text{(a) AutoMDT concurrency after offline training}
%   \end{minipage}
%   % \hfill
%   \begin{minipage}[b]{0.4\textwidth}
%     \centering
%     \includegraphics[width=\textwidth]{images/concurrency_ft_vs_wo_AutoMDT_FT_network_bn.pdf}
%     \vspace{0.5em}
%     \text{(b) AutoMDT concurrency after finetuning}
%   \end{minipage}
  
%   \caption{Difference in predicted concurrency by the two models. One is trained offline, while the other undergoes an additional online fine-tuning phase. Their performance is nearly identical.}
%   \label{fig:finetune}
% \end{figure*}

\subsection{Online Fine-tuning}
As the entire training process was conducted offline, we experimented with online fine-tuning to verify that our simulator produced the intended results. For this purpose, we used a model obtained from offline training and further trained it online for 120 episodes (2 hours). The performance of the fine-tuned model is very close to the offline-trained model. In numerical analysis, we observed that the fine-tuned model used 1\% less concurrency while achieving the same transfer speed. Due to this negligible improvement, we decided to exclude online fine-tuning from our proposed solution.

% As illustrated in Figure~\ref{fig:finetune},

% \begin{figure}[ht]
%     \centering
%     \includegraphics[width=0.45\textwidth]{images/sota_comparison.pdf}
%     \caption{Performance evaluation of AutoMDT.}
%     \label{fig:sota}
% \end{figure}

% \subsection{Comparison to State-of-The-Art}
% We compared AutoMDT with commercially most widely used Globus, Falcon \cite{arifuzzaman2023falcontpds}, and State-of-The-Art Marlin. When using Globus, with the open source GCT 6.2 GridFTP, we used the values of concurrency and parallelism as 4 and 8 respectively. In case of Falcon, Marlin, and AutoMDT the concurrency is adaptive. We compared them in two scenarios as illustrated in Figure~\ref{fig:sota}. The scenarios are the same small files dataset and large files dataset as described earlier. In both of the cases, AutoMDT shown superior performance. In small files dataset, AutoMDT's speed were found to be 13.9x, 5.7x, and 1.7x of the speeds achieved by Globus, Falcon and Marlin respectively. On the other hand, for large files dataset, AutoMDT achieved 6.6x, 2.7x, 1.3x speed compared to Globus, Falcon and Marlin respectively. This result showcases AutoMDT's superiority of optimizing available resources.

\begin{table}[h]
    \centering
    \caption{End‑to‑end Transfer Speed Comparison}
    \label{tab:thrpt}
    % for three representative datasets
    \begin{tabular}{lcccccc}
        \toprule
        \textbf{Dataset} & \textbf{Total Size} & \textbf{Globus} & \textbf{Marlin} & \textbf{AutoMDT} \\
        \midrule
        % A (Small) & 20 GB & 1\,328.9 & 11\,162.4 & 18\,547.9 \\
        A (Large) & 1 TB  & 3,652.2 & 18,066.8 & 23,988.0 \\
        B (Mixed) & 1 TB  & 2,325.9 & 13,721.5 & 16,915.8 \\
        \bottomrule
    \end{tabular}
\end{table}

\subsection{Comparison with State-of-the-Arts}
We periodically perform data transfers in FABRIC Testbed (NCSA to TACC) using Globus, Marlin, and \name~for both large and mixed datasets. The experiments were repeated several times each day for a week, and all results in Table~\ref{tab:thrpt} represent the averages of those runs. For Globus, we initially used globus-online; however, we found that the transfer speed was unusually slow, even with integrity verification disabled. As a result, we used globus-url-copy from the open-source Grid Community Toolkit (GCT 6.2) for these experiments. Globus relies on heuristic and static configurations and cannot adapt to changing network conditions. We set the concurrency to 4 and parallelism to 8. In contrast, both Marlin and \name~employ online optimization to dynamically adjust concurrency levels as needed. As shown in Table~\ref{tab:thrpt}, \name~significantly outperforms both Marlin and Globus. 
For the large-file set (Dataset A), it reaches 23.9 Gbps, which is $6.57X$ and $1.33X$ the speed of Globus and Marlin. For the mixed-file set, it reaches 16.9 Gbps, which is $7.28X$ and $1.23X$ faster than the same baselines. 
These results demonstrate that \name~utilizes I/O and network resources more efficiently than current state-of-the-art tools. While it may be somewhat unfair to compare Globus's static configuration with our adaptive approach, our primary objective is to demonstrate that \name~can dynamically scale in response to available resources. In contrast, Globus’s preset parameter values often lead to underutilization of available bandwidth, as system administrators typically avoid aggressive settings to minimize system overhead and prevent network congestion. Moreover, given the constantly evolving nature of systems and network conditions, fixed parameters are likely to be suboptimal for most of the transfer duration. 
% Falcon~\cite{arifuzzaman2023falcontpds},  
% and Marlin~\cite{marlin}.  
% Falcon, Marlin, and AutoMDT pick their thread counts on the fly; Globus uses the fixed settings above. The two workloads (small and large) are the same as in the earlier sections.

% Figure~\ref{fig:sota} shows the results.  AutoMDT is fastest in both cases.  
% For the small‑file set it reaches {18.5}{Gbps}, which is 
% \textbf{13.9×}, \textbf{5.7×}, and \textbf{1.7×} the speed of Globus, Falcon, and Marlin.  
% For the large‑file set it reaches {24.0}{Gbps}—\textbf{6.6×}, \textbf{2.7×}, and \textbf{1.3×} faster than the same baselines.  
% These results show that AutoMDT uses I/O and network resources more efficiently than current state‑of‑the‑art tools.
%\vspace{-2mm}

\section{Conclusion}\label{conc}
In today's era of high-performance computing, data transfer optimization is essential to fully utilize high-speed networks. Although previous works have addressed this issue, they suffer from two major challenges in modern HPC infrastructures. First, most solutions follow a monolithic architecture that uses the same concurrency for all components, leading to significant system resources overhead and suboptimal transfer throughput. Second, the existing modular architecture-based solutions fail to account for the memory buffer dynamics at both the sender and receiver ends leading to unstable and suboptimal performance. In this work, we demonstrated how deep reinforcement learning can efficiently solve this problem. The DRL agent learns the systems dynamics offline with the help of a testbed simulator and the optimizer can reach the optimal concurrency settings up to $8X$ faster. Experimental evaluations on several testbeds show that the agent successfully identifies near-optimal solutions, achieving up-to $68\%$ faster transfer completion times compared to the state-of-the-art solutions.
%\vspace{-2mm}

\bibliographystyle{plain}
\bibliography{references}

\end{document}